\documentclass[10pt,sigconf]{acmart}
\usepackage{graphicx}
\usepackage{balance}

\usepackage{cleveref}
\usepackage[inline]{enumitem}

\usepackage{url}

\usepackage[utf8]{inputenc}
\usepackage{booktabs}
\usepackage{subcaption}

\usepackage{mathtools}
\DeclarePairedDelimiter\ceil{\lceil}{\rceil}

\usepackage{diagbox}
\usepackage{multirow}
\usepackage{arydshln}

\usepackage{tikz}
\usepackage{pgfplots}
\usepackage{listings}
\pgfplotsset{compat=newest}

\newcommand{\RebelRig}{Sys1\$}
\newcommand{\SIM}{Sys2\$}
\newcommand{\HPC}{Sys10\$}

\begin{document}


\title{Training for Speech Recognition on Co-processors}

\author{Sebastian Baunsgaard}
\email{sebab@itu.dk}
\affiliation{
\institution{ITU}
}

\author{Sebastian B. Wrede}
\email{sbwr@itu.dk}
\affiliation{
\institution{ITU}
}

\author{Pınar Tözün}
\email{pito@itu.dk}
\affiliation{
\institution{ITU}
}

\begin{abstract}

  Automatic Speech Recognition (ASR) has increased in popularity in recent years.
  The evolution of processor and storage technologies has enabled more advanced ASR mechanisms,
  fueling the development of virtual assistants such as Amazon Alexa, Apple Siri, Microsoft Cortana, and Google Home.
  The interest in such assistants, in turn, has amplified the novel developments in ASR research.
  
  However, despite this popularity, there has not been a detailed training efficiency analysis of modern ASR systems.
  This mainly stems from:
  the proprietary nature of many modern applications that depend on ASR, like the ones listed above;
  the relatively expensive co-processor hardware that is used to accelerate ASR by big vendors to enable such applications;
  and the absence of well-established benchmarks.
  The goal of this paper is to address the latter two of these challenges.

  The paper first describes an ASR model, based on a deep neural network inspired by recent work in this domain, and our experiences building it.
  Then we evaluate this model on three CPU-GPU co-processor platforms that represent different budget categories.
  Our results demonstrate that utilizing hardware acceleration yields good results even without high-end equipment.
  While the most expensive platform (10X price of the least expensive one) converges to the initial accuracy target
  10-30\% and 60-70\% faster than the other two,
  the differences among the platforms almost disappear at slightly higher accuracy targets.
  In addition, our results further highlight both the difficulty of evaluating ASR systems
  due to the complex, long, and resource intensive nature of the model training in this domain,
  and the importance of establishing benchmarks for ASR.

\end{abstract}

\maketitle

\section{Introduction}
\label{sec:introduction}

Automatic Speech Recognition (ASR) has been an active research area for decades \cite{ASRBook, MFCCOrg1980, ASRConversationalSound, ASRDigits1952}.
Its popularity and complexity keep increasing as a result of the popularity of various virtual assistants \cite{Alexa, Siri, Cortana, GoogleAssistant}.
Earlier approaches to ASR were based on statistical models such as the Gaussian Mixture Model - Hidden Markov Model (GMM-HMM) \cite[p. 19]{ASRBook}.
However, in recent years, the emergence of neural networks has also influenced ASR
\cite{DeepSpeech1, DeepSpeech2, NeuralTransducers}.

The computational requirements for the training and inference of such neural networks are immense \cite{GoldenAge,BenchmarkingSOTADL}. 
Most calculations in neural network models are independent matrix operations.
Therefore, they are a natural fit for hardware acceleration
on specialized multithreaded hardware such as GPUs \cite{OriginalGPUDNN, OriginalGPUDNNFollowUp}.
Any performance improvement through such hardware acceleration is significant for researchers or developers trying to enhance their machine learning models
as they have to experiment with several different hyperparameters before deciding on the final set of hyperparameters.

To determine better software and hardware setups for any application domain, benchmarks are essential.
Despite its popularity, ASR is under-studied when it comes to established benchmarks and benchmarking.
There have been benchmarking studies of different machine learning and deep learning platforms \cite{liu2018benchmarking, BenchmarkingSOTADL} without focusing on ASR.
A prominent benchmarking framework for machine learning is MLPerf \cite{mlperf}, which has a benchmark for ASR under development that has not been released yet.
ASR differs from these existing machine learning and neural network benchmark domains
due to its requirement of a more complex model training and larger space consumption of the dataset and the model. 
QuTiBench \cite{xilinxNNBench} and DeepBench \cite{deepbench} are recent works that also incorporate benchmarking the training of deep neural networks for ASR
in addition to the other domains.
They are complementary to this paper as
the former is still quite new and in its early stages, and the latter focuses more on individual operations rather than more complex end-to-end model training.

\begin{figure}
  \centering
  \includegraphics[width=0.9\linewidth]{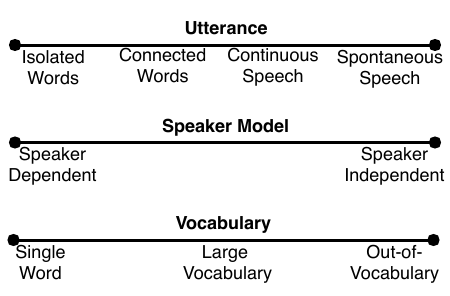}
  \caption{Speech properties.}
  \label{fig:SpeechProperties}
\end{figure}

Our goal in this paper, instead, is to have a more in-depth understanding of ASR 
by focusing on the ASR task of converting speech to text and utilization of CPU-GPU co-processors while training a neural network model for this task.
Our contributions are as follows:
\begin{list}{\labelitemi}{\leftmargin=1.5em}
\item{We discuss our experience with building a well-established acoustic model that converts speech to text based on a deep neural network inspired by recent work
  \cite{DeepSpeech1, DeepSpeech2, NeuralTransducers}.
We use the training of this model as a benchmarking tool while analyzing acoustic model training behavior on different types of hardware platforms.}
\item{We evaluate three CPU-GPU co-processor platforms that represent three budget categories for training this model:
  a low-budget platform built by us via repurposing a crypto mining rig,
  and two more expensive platforms that represent different generations of commodity co-processor server hardware
  (a recent previous generation and a modern high-end one).}
\item{We take Time-to-Accuracy (TTA) as the primary metric in this evaluation in order to emphasize both training efficiency and model accuracy.
  The results demonstrate that utilizing hardware acceleration yields good results even without high-end co-processors.
  The most expensive platform (10X price of the least expensive one) converges to the initial accuracy target
  10-30\% faster than the platform that is 2X the price of least expensive one and 60-70\% faster than the least expensive platform.
  However, the differences among the platforms almost disappear at slightly higher accuracy targets,
  which take roughly a couple of days to reach.}
\end{list}
Our overall experience also highlights that
it is extremely difficult to establish ASR models that both
perform well in terms of accuracy and are hardware-conscious while training.
While co-processors are becoming more widely available to everyone, it
increases the opportunities for accelerating
not only ASR but also other neural network training. 
With this development, how to better utilize the co-processors to get the best price/performance ratio is unclear.
Our results emphasize that this ratio can indeed be very poor
in practice for acoustic model training.
Therefore,
it is highly important to have well-established benchmarks in this domain and
other similarly complex domains to be able to characterize the performance behavior of both high-end and low-budget platforms.

The rest of the paper is organized as follows.
\Cref{sec:background} introduces terminology related to ASR and gives an overview of our end-to-end system that converts speech to text.\footnote{We plan to open-source once the paper is accepted.}
\Cref{sec:related_work} surveys related work that inspired this system and
recent work on benchmarking machine learning models.
\Cref{sec:ExperimentalMethodology} presents the experimental setup, including model parameters and methodology,
as well as the implementation details for efficient model training using the TensorFlow framework.
\Cref{sec:results} discusses the results of the experimental evaluation on three co-processors.
Finally, \Cref{sec:conclusion} provides a summary of the paper,
discussion of results, and conclusion.

\section{Speech-to-Text}
\label{sec:background}

Speech has three main properties that express the difficulty of converting it to text
as defined in \cite{VimalaCSpeechRecognition} and illustrated by \Cref{fig:SpeechProperties}.
\textbf{\textit{Types of speech utterance}} characterizes the pauses between the words and the coherence of the utterances.
It goes from \textit{isolated words}, which has only a single word at a time with long pauses between the words,
to \textit{spontaneous speech}, which is natural speech with false-starts, mispronunciations, and no pauses between the words. 
\textbf{\textit{Speaker model}} characterizes the number and variation of speakers and the setting of the recording.
It goes from \textit{speaker dependent}, where a specific speaker and a specific setting are required,
to \textit{speaker independent}, where a large number of different speakers are recorded in a wide range of different settings. 
\textbf{\textit{Vocabulary}} characterizes the number of different words and phrases.
It goes from a \textit{single word} to \textit{out-of-vocabulary}, which includes unknown words.
The further to the right in \Cref{fig:SpeechProperties} a speech dataset is, the more difficult it is to convert the speech to text. 
The dataset chosen in this paper falls toward the right-hand side of this scale as described in \Cref{sec:ExperimentalMethodology:Training_Data}.
In addition to the speech properties, the recording can be more or less noisy. Noise is all the sounds captured in the recording that is not created by the voice speaking. 

\begin{figure}
    \centering
    \includegraphics[width=\linewidth]{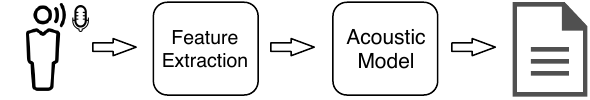}
    \caption{Process of converting speech to text.}
    \label{fig:process}
\end{figure}

The process of converting speech to text is composed of two broad components as \Cref{fig:process} illustrates \cite[p. 4]{ASRBook}:
\emph{Feature Extraction} (FE) and \emph{Acoustic Model} (AM).
Extracting features from audio in FE before training the acoustic model enables performance optimizations in AM.
It is also common to add a \emph{Language Model} (LM) trained separately on text data to improve the output of AM \cite{DeepSpeech1, DeepSpeech2, NeuralTransducers, ListenAttendSpell, newLAS_ICASSP, InterSpeechStateOfTheArtPerformanceLibrispeech}.
Among these components, AM takes the longest time in an end-to-end system for ASR that converts speech to text since it contains the training of the machine learning model.
On the hardware platforms used in this paper (\Cref{sec:ExperimentalMethodology:hw}), FE takes a few hours whereas training the AM takes several days.
Therefore, this paper focuses on AM when benchmarking the different hardware platforms.
The rest of this section provides details about FE and AM components in our system.

\subsection{Feature Extraction}
\label{sec:background:FE}

\begin{figure}
    \centering
    \includegraphics[width=\linewidth]{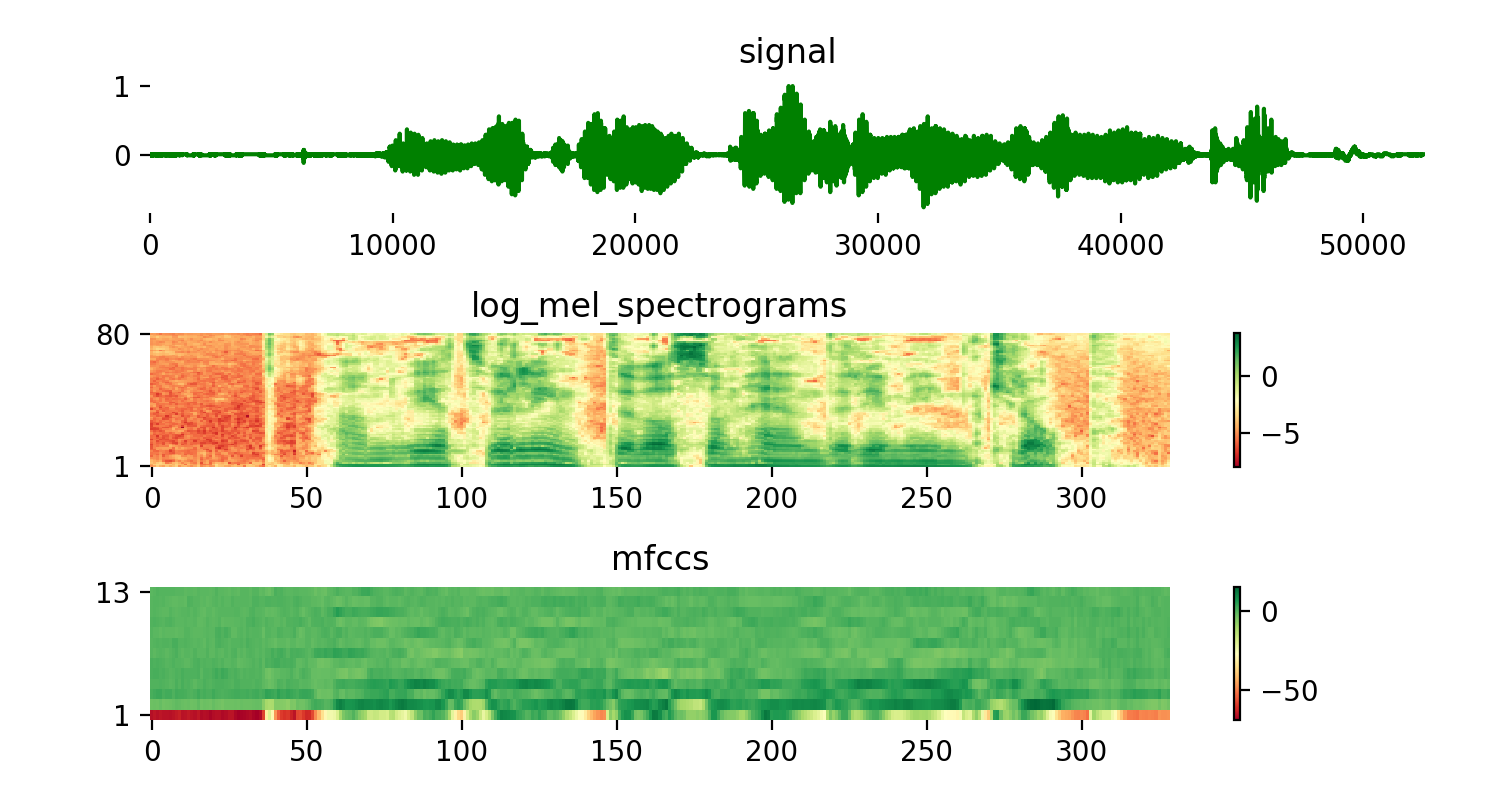}
    \caption{Example of FE process converting audio signal to Log Mel and MFCC.}
    \label{img:FE_steps_visualized}
\end{figure}

The Feature Extraction (FE) takes sound as input, extracts important information from audio, and returns windows of features. 
The windows are also sometimes called frames, and these are sampled based on window width and stride size.
A typical starting feature is calculated using a Short-Time Fourier Transform (STFT) on the framed parts of a sound file \cite[p. 41]{MusicSimilarityandRetrival}. 
The STFT takes the sound from the \emph{Time Domain} to the \emph{Frequency Domain} \cite[p. 16]{HumanandMachineHearingLyon}.
The Time Domain represents sound as an amplitude over time,
whereas the Frequency Domain represents sound as intensities over different frequencies for each frame of the sound. 
The STFT returns a \emph{magnitude spectrogram}, which is divided into Hertz frequencies.
This representation can be converted to Mel-scale giving us the features referred to as \emph{Mel frequencies} \cite[pp. 53 - 56]{MusicSimilarityandRetrival}. 
The logarithm of the Mel frequencies produces \emph{Log Mel} frequencies (middle plot in \Cref{img:FE_steps_visualized}).
Finally, the Log Mel Frequencies can be used to calculate the \emph{Mel-Frequency Cepstral Coefficients} (MFCC) by applying a Discrete Cosine Transform (DCT)
\cite[p. 55]{MusicSimilarityandRetrival} (bottom plot in \Cref{img:FE_steps_visualized}).

FE can happen in different ways even if the same type of feature extraction is performed
because of the different ways the windows can be sampled and the number of features extracted from each window. 
Depending on the parameters chosen, the features have different levels of details. 
For instance, Chiu et al. \cite{uniLAS} use $80$ Log Mel features with $25ms$ windows and $10ms$ stride,
whereas Chan et al. \cite{ListenAttendSpell} use $40$ Log Mel features with only $10ms$ windows. 
In this paper, $80$ Log Mel features are combined with $13$ MFCC features with a $10ms$ stride.
We pick a window size of $512$ samples corresponding to $32ms$ with a sampling frequency of $16kHz$.
There are many more features that could be explored \cite[pp. 40 - 50]{MusicSimilarityandRetrival}.
The features chosen in this paper and their calculated values are based on a combination of
an empirical analysis of the dataset and 
the wisdom learned from related work
\cite{coldfusion, uniLAS, DeepSpeech1, ListenAttendSpell, CocktailPartyProblem_MultiSpeaker, IBM_MALACH_corpus}.

\subsection{Acoustic Model}
\label{sec:background:AM}

\begin{figure}
    \centering
    \includegraphics[width=0.45\linewidth]{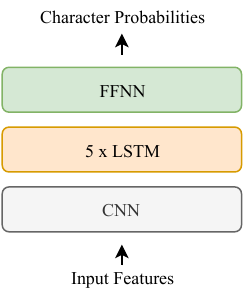}
    \caption{Overview of the layers of our AM.}
    \label{fig:model}
\end{figure}

The Acoustic Model (AM) takes the features from FE as input and generates a probability distribution over the dictionary \cite[p. 4]{ASRBook}.
In recent years, AM has been based on neural networks and several combinations of different neural network layers have been explored
\cite{ASRBook, DeepSpeech1, DeepSpeech2, ListenAttendSpell, NeuralTransducers}. 
The commonly used layers are Feed-Forward Neural Network (FFNN), Convolutional Neural Network (CNN), and Long Short-Term Memory (LSTM). 
The model used in this paper is inspired by Baidu Research's Deep Speech 2 system \cite{DeepSpeech2},
which uses 1-3 CNNs as the first layers followed by 1-7 LSTM layers, and ending with a single FFNN.
\Cref{fig:model} gives an overview of the model this paper uses,
where the number of CNN, LSTM, and FNNN layers are 1, 5, and 1, respectively that is the chosen settings for this paper. 
The following subsections first describe these layers and then other important components of the acoustic model we built.

\subsubsection{Neural Network Layers}
An \textbf{\textit{FFNN}} can be seen as a directed acyclic graph, where the nodes are functions and the output of each function is a directed edge.
The nodes are organized in layers, where each layer represents a matrix multiplication of the input with a weight matrix for that layer.
This weight matrix is also called a \textit{kernel}.
The FFNN we use in this paper is based on \texttt{tf.contrib.slim.fully\_connected} from TensorFlow.

A \textbf{\textit{CNN}} slides a kernel over the input doing matrix multiplications for each step.
A CNN depends on the kernel \textit{width}, which is the number of input elements included in each computation step,
and \textit{stride}, which is the number of inputs the kernel is slided over between each step.
A CNN can be used to do computations on values from different time steps.
The convolution used in this paper is \texttt{tf.nn.conv1d} from TensorFlow,
which conducts a one-dimensional convolution on a three-dimensional input.
This means that a convolution is done on the time-dimension of an input consisting of features, time, and batch. 

An \textbf{\textit{LSTM}} applies a function to all elements of an input sequence of arbitrary length while transferring information from one time-step to the next.
This property makes it ideal for processing time series data such as audio.
The size of an LSTM refers to the number of hidden states transferred between the time-steps.
Increasing the size of an LSTM increases the capacity of the internal state, which improves the model's ability to fit to the training dataset. 
The LSTM in this paper is based on \texttt{tf.nn.rnn\_cell.LSTMCell}, combined with \texttt{tf.nn.dynamic\_rnn} from TensorFlow.
TensorFlow also has a GPU-optimized LSTM version called \texttt{tf.contrib.cudnn\_rnn.CudnnLSTM},
but this implementation did not improve performance when adding the activation function and batch normalizations between LSTM layers. 

The weights of the layers are initialized using Xavier Initialization, which scales all weights to a uniform distribution within a range \cite{XavierInitialization}. 
The Xavier Initialization is done using \texttt{tf.contrib.layers.xavier\_initializer} from TensorFlow since it achieved good results.

\subsubsection{CTC Loss \& Beam Search}
\label{sec:background:AM:ctcbeam}

The output of the FFNN is processed differently during training and evaluation.
During training, a \textit{loss} value has to be calculated for each element of the batch to determine how to fit the model to the training data. 
During evaluation, the model has already been adapted to the training data and the output of the FFNN should instead be converted to specific characters representing the output sentence of the model. 

The loss function used depends on the problem domain.
In ASR a common function is \textit{Connectionist Temporal Classification} (CTC) \cite{CTC, DeepLearningGoodfellow, DeepSpeech2, NeuralTransducers}. 
CTC converts the output of the FFNN to probabilities over the alphabet and aligns it to the label sequence. 
Given the input sequence of probabilities generated by the CTC denoted $Y$ and the label sequence $l$,
the \textbf{\textit{CTC loss}} is the sum of the probability of different alignments of $l$ in the probability sequence $Y$. 

When evaluating a model, the output probabilities have to be evaluated efficiently to produce the characters of the output sentence.
This is done with \textbf{\textit{Beam Search}}, which is a modified Breadth First Search (BFS) with a specified search width that limits the search through the tree.
Beam Search is not guaranteed to find an optimal solution, but it is more efficient than BFS.
The efficiency depends on the search width. 
Beam Search can be improved by combining it with a Prefix Tree and a Language Model as presented by Scheidl et al. \cite{Beam_Language_Github, beam_language_model_LM}.
This version of Beam Search constrains its search to words in a dictionary while simultaneously allowing arbitrary non-word characters between words in the dictionary.
In this paper, we adopt the implementation from Scheidl et al. and build the dictionary from the words in our training dataset.
This limits the variation of words in the dictionary compared to the alternative which is to build the dictionary from a separate dataset with a greater variation of words. 
This alternative has been done by related work and relevant datasets for building a dictionary have been provided by LibriSpeech \cite{LibriSpeech}. 

\subsubsection{Optimization Algorithm}
An \textbf{\textit{optimization algorithm}} in machine learning defines how to adapt the model weights to reduce the loss of the model for the given training set. 
We use \textit{AdaDelta} \cite{adadelta} which is able to both increase and decrease the learning rate during training.
This property also implicitly removes the need of specifying initial learning rates.
AdaDelta is also used by Bahdanau et al. \cite{attention-based-ASR}. 
The implementation of AdaDelta is from \texttt{tf.train.AdadeltaOptimizer} in TensorFlow.

\subsubsection{Regularization}
A model needs to not only fit to the training dataset, but also generalize to previously unseen data. 
The error calculated on the training set is \emph{training error} and
the error calculated on the test set is \emph{test error} or \emph{generalization error}.
\textbf{\textit{Regularization}} aims to reduce the gap between the training and test error \cite[pp. 107-108]{DeepLearningGoodfellow}. 
One of the most important regularization strategies is dropout,
where some of the units in the neural network are randomly dropped during training \cite{DropoutAlex}.
We use this strategy in all of the layers of our model since it reduces the test error which improves the test accuracy of the model.

\subsubsection{Batch Normalization}
\textbf{\textit{Batch Normalization}} is applied between each layer of the model \cite{BatchNormalizing} \cite[pp. 313-317]{DeepLearningGoodfellow}.
This technique helps the layers train more efficiently and independently of each other.

A general normalization is done with the following equation \cite{BatchNormalizing}: 

\begin{equation}
    \widehat{x}_i^{(k)} = \frac{x_i^{(k)} - \mu_{\beta}^{(k)}}{\sqrt{{\sigma^{2}}_{\beta}^{(k)} + \delta}}
\end{equation}

$x_i^{(k)}$ is dimension $k$ of input element $i$ in batch $\beta$. 
The mean of dimension $k$ of all elements in batch $\beta$ is denoted by $\mu_{\beta}^{(k)}$ and the variance of dimension $k$ in batch $\beta$ is ${\sigma^{2}}_{\beta}^{(k)}$. 
To prevent division by zero in the case where the variance equals zero, the small non-zero value $\delta$ is added in the denominator.
The normalized value of $x_i^{(k)}$ is denoted $\widehat{x}_i^{(k)}$. 
The normalized dimensions will have a mean of zero and a variance of approximately 1 \cite{BatchNormalizing}.

\section{Related Work}
\label{sec:related_work}

We mention related work throughout the paper wherever it is necessary.
This section, in particular, details the work that inspired the system described in
\Cref{sec:background:AM}, and
recent efforts on establishing benchmarks and analyzing machine learning models.

\subsection{Speech to Text Models Based on Neural Networks}
\label{sec:related_work:nn}

Baidu Research's Silicon Valley AI Lab is a large research group that
among other fields also does research
on building end-to-end deep learning models for speech recognition
\cite{DeepSpeech1, DeepSpeech2, NeuralTransducers}.
Their earlier major work, Deep Speech 1 \cite{DeepSpeech1},
is one of the preliminary works that uses neural networks
for the whole speech to text learning pipeline in contrast to prior work
that used neural networks in a limited part of the whole pipeline.
Deep Speech 1 is also the foundation of Mozilla's open-source TensorFlow speech recognition implementation \cite{DeepSpeechGithub}. 

Deep Speech 2 \cite{DeepSpeech2} is a followup to Deep Speech 1.
It experiments with several convolution layers and up to 7 layers of LSTM and bidirectional LSTM.
Furthermore,
it introduces GPU-optimized implementations of CTC loss and
efficient gradient sharing among GPUs,
which improves latency and throughput.
However,
the impact of these optimizations on the training time are not based on TTA,
but on the time to go through an epoch and the time spent on individual operations
such as their implementation of the all-reduce algorithm.
This approach to evaluating performance has been criticized in DawnBench
\cite{DawnBench},
because it does not make a strong link across 
end-to-end training efficiency, hardware utilization, and statistical performance. 

Another followup work from the same research group \cite{NeuralTransducers}
investigates the impact of different transducers on the ASR systems.
The models presented in Deep Speech 1 \& 2 are based on CTC,
and this work compares these approaches to an RNN-Transducer and an attention model.
The conclusion is that both RNN-Transducers and attention models outperform
the CTC-based model if the models are allowed to look at the entire input
(meaning both forward and backward model parts).
The paper also shows that the CTC forward-only models have better results than
their forward-only RNN-transducer and attention models.
We choose to work with a CTC-based model instead of
RNN-Transducers and attention models in order to avoid the need for
the entire input for one classification.

While the main inspiration for our acoustic model comes from the work of Baidu Research mentioned above,
there have been other proposals for neural network models for speech recognition,
which we also take influence from,
such as LAS from Google \cite{ListenAttendSpell, newLAS_ICASSP}.

\subsection{Benchmarking Machine Learning Models}
\label{sec:related_work:benchmarks}

There have been several benchmarking studies in the recent years focusing on machine learning and deep learning.
Shi et al. \cite{BenchmarkingSOTADL} analyze popular deep learning frameworks, such as TensorFlow and Torch.
They take training time per mini-batch as the main metric.
As mentioned earlier, this is problematic as it does not account for the total training time of the model in different frameworks.
DawnBench \cite{DawnBench} treats training time per mini-batch as a \textit{proxy metric} rather than a main one.
Instead, DawnBench measures end-to-end performance of training and inference.
It focuses on the time to accuracy and throughput of models as mini-batch size, optimization algorithm, number of GPUs, etc. varies.
These metrics are also adopted by MLPerf \cite{mlperf},
which is the most popular benchmarking framework for machine learning today.
Liu et al. \cite{liu2018benchmarking} also adopts the metrics from DawnBench and
experiment with image recognition models with the default configurations of the different deep learning frameworks.
However, none of these works have focused on ASR so far.

MLPerf has plans for releasing a benchmark for ASR in the near future.
QuTiBench \cite{xilinxNNBench} and DeepBench \cite{deepbench}
describe benchmarks for different deep neural network training domains
including ASR.
DeepBench focuses on individual operations rather than end-to-end training,
while QuTiBench focuses on end-to-end training.
We focus on end-to-end training as well and
specifically focus on its performance on CPU-GPU co-processors.

\section{Experimental Methodology}
\label{sec:ExperimentalMethodology}

There are different dimensions to analyze when it comes to any benchmarking and workload characterization study.
Our goal in this paper is to analyze the behavior of training a state-of-the-art acoustic model for ASR on different types of co-processor hardware.
We focus on the acoustic model as its training is the major component in an end-to-end system for ASR (as \Cref{sec:background} also mentions).
We target CPU-GPU co-processors as training of neural network models is a natural fit for hardware acceleration on GPUs
because of the independent matrix operations that are embarrassingly parallel
and such co-processors are heavily utilized for this purpose in the cloud today
\cite{GoldenAge, OriginalGPUDNN, OriginalGPUDNNFollowUp}.

In our evaluation, we would like to answer the following couple of questions:
\begin{list}{\labelitemi}{\leftmargin=1.5em}
\item{Does one depend on expensive high-end co-processor infrastructure for efficient ASR?}
\item{How does the training of an established acoustic model utilize CPU-GPU co-processors?}
\end{list}

To be able to answer these questions,
we choose Time-to-Accuracy (TTA) as our main metric (\Cref{sec:ExperimentalMethodology:metrics}),
establish three CPU-GPU co-processor platforms that represent different budget categories (\Cref{sec:ExperimentalMethodology:hw}),
use a dataset that has difficult characteristics based on the scale in \Cref{fig:SpeechProperties} (\Cref{sec:ExperimentalMethodology:Training_Data}),
and optimize the training parameters and setup on the co-processors being used to be fair to each hardware platform and avoid misleading conclusions (\Cref{sec:ExperimentalMethodology:training}).

\subsection{Metrics}
\label{sec:ExperimentalMethodology:metrics}

We focus on several metrics in our evaluation: \emph{accuracy}, \emph{time-to-accuracy}, \emph{throughput}, and \emph{utilization}. 
These metrics are chosen to assess different aspects of the acoustic model and hardware platforms. 
We pick \emph{time-to-accuracy} as the primary metric since it covers both the accuracy and efficiency of training. 
The following subsections describe each metric in further detail.

\subsubsection{Accuracy}
\label{sec:ExperimentalMethodology:metrics:accuracy}

As \Cref{sec:background} explains, the ASR system we build aims at converting speech to text.
Therefore, the \emph{input} while evaluating the trained acoustic model is sound files with speech
and the \emph{output} is the text version of what is being said in those sound files.
The accuracy of this output is measured in Word Error Rate (WER) and Character Error Rate (CER).

\begin{equation} \label{eq:wer}
    WER =  \frac{e(x,y)}{W} = \frac{I + D + S}{W}
\end{equation}

As \Cref{eq:wer} shows, WER definition is based on \emph{edit} distance or, more specifically, \emph{Levenshtein} distance $e(x,y)$.
It is the sum of insertions $I$, deletions $D$, and substitutions $S$ needed to convert the output sentence $x$ to the target sentence $y$.
An \emph{insertion} is a word added to the output. 
A \emph{deletion} is a word removed from the output. 
A \emph{substitution} is a word replaced with another word.
Finally, $W$ refers to the number of words in the target sentence $y$.
We use the Python package \emph{Jiwer} \cite{wer-asr-python} to calculate WER in this paper. 

\begin{equation} \label{eq:cer}
    CER = \frac{e(x,y)}{\operatorname{max}(|x|,|y|)}
\end{equation}

As \Cref{eq:cer} shows, the definition of CER is similar to WER.
The only difference is that the denominator is the maximum number of characters in either the output sentence $x$ or the target sentence $y$.

\subsubsection{Time-to-Accuracy}

Time-to-Accuracy (TTA) is the primary metric in this evaluation
in order to emphasize training of a model with high accuracy in a short amount of time
(as also used by \cite{DAWNBenchTTA, Crossbow}).
TTA is the time it takes to train a model to a target median accuracy over the last $t$ epochs.
An \emph{epoch} is an iteration of the entire \emph{training} dataset during the training phase of the model.
The accuracy of a single epoch is the accuracy (as defined by \Cref{sec:ExperimentalMethodology:metrics:accuracy})
of a single iteration over the entire \emph{validation} dataset during the evaluation phase of the model
after that particular epoch.
The \emph{median accuracy} is the middle value in the sorted list of accuracies from the last $t$ epochs.
The reason for considering the accuracy of several epochs in TTA is to make sure that the accuracy reached is persistent.

To calculate $TTA(a,e)$, where $a$ is the median accuracy target and $e$ is the number of epochs, $a$ and $e$ needs to be determined. 
We set $e$ to three and $a$ to different values depending on the specific accuracy metric used (WER or CER).
The $TTA(a,e)$ is measured in minutes from the creation of the model until the accuracy target is reached.

\subsubsection{Throughput}

The throughput is defined as elements processed per second $E_{ps}$. 
If $t$ is the time it takes for a batch to be processed in seconds,
and $b$ is the number of elements processed in that batch,
then elements per second is:   

\begin{equation} \label{eq:throughput}
    E_{ps} = \frac{b}{t}
\end{equation}

\subsubsection{Utilization}

The last metric is the utilization of the processors,
where we measure the active usage of CPUs and GPUs while training the model.

The CPU utilization is measured based on the average percentage utilization over 0.03 seconds using the \emph{top} command of linux:
\begin{lstlisting}[language=bash]
  top -n 1 -d 0.03 -p <process_id>
\end{lstlisting}

The GPU utilization is measured through \emph{The NVIDIA System Management Interface (nvidia-smi)} \cite{NVIDIA_system_management_interface}:
\begin{lstlisting}[language=bash]
  nvidia-smi --format=csv,noheader \
       --query-gpu=utilization.gpu
\end{lstlisting}

\subsection{Hardware}
\label{sec:ExperimentalMethodology:hw}

\begin{figure*}
    \begin{subfigure}{0.33\linewidth}
        \centering
        \includegraphics[width=\linewidth]{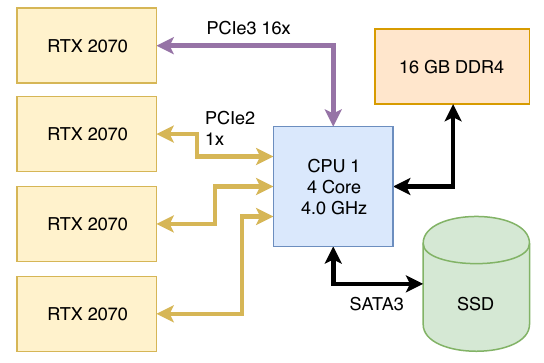}
        \subcaption{\RebelRig}
        \label{fig:RebelRig_Architecture}
    \end{subfigure}
    \begin{subfigure}{0.33\linewidth}
        \centering
        \includegraphics[width=\linewidth]{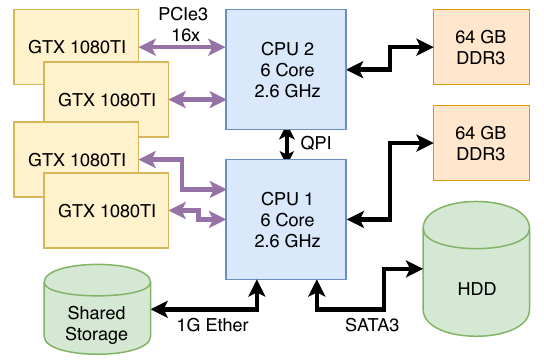}
        \subcaption{\SIM}
        \label{fig:Sim_Architecture}
    \end{subfigure}
    \begin{subfigure}{0.33\linewidth}
        \centering
        \includegraphics[width=\linewidth]{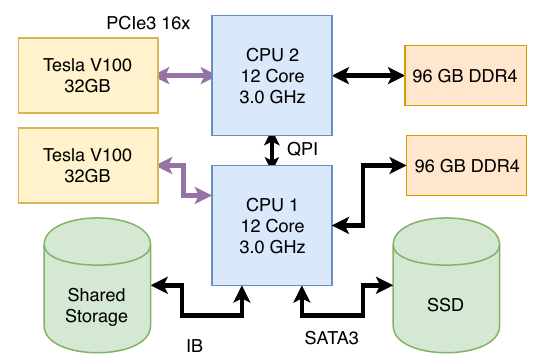}
        \subcaption{\HPC}
        \label{fig:HPC_Architecture}
    \end{subfigure}
    \vspace{-0.4cm}
    \caption{Overview of the co-processors used (left to right, from least expensive to most expensive).}
    \label{fig:hw_architectures}
\end{figure*}

\begin{table}
    \centering
    \begin{tabular}{@{ \ }llr@{ \ }}
    \toprule
    System        & Description           & Cost      \\
    \midrule
    \RebelRig      & Total                 &  2'605.29\$           \\
                   & GPUs only             &  1'980.00\$           \\
                   & CPU  only             &    354.25\$           \\
    \midrule
    \SIM           & Total                 &  5'699.95\$           \\
                   & GPUs only             &  3'599.97\$           \\
                   & CPUs only             &     97.98\$           \\
    \midrule
    \HPC           & Total                 & 25'999.00\$           \\
                   & GPUs only             & 17'956.00\$           \\
                   & CPUs only             &  5'422.20\$           \\
    \bottomrule
    \end{tabular}
    \caption{
        Costs of hardware platforms estimated based on prices from amazon.com in October 2019.
        The total cost is composed of CPU, GPU, RAM, PCIe riser card, chassis, and motherboard costs
        (not including storage).
        }
    \label{tab:SystemCosts}
\end{table}

We use three hardware platforms for this study: \RebelRig, \SIM, \HPC.
They represent three budget categories in terms of co-processor hardware.
\Cref{tab:SystemCosts} gives the cost breakdown for these co-processors,
\Cref{fig:hw_architectures} illustrates their topology, and
the following subsections describe them in more detail.
We name them based on their relative total costs, i.e., \SIM~and \HPC~are roughly 2X and 10X the cost of \RebelRig, respectively. 

\subsubsection{\RebelRig}
\label{sec:ExperimentalMethodology:hw:rebel}
The first hardware platform is \textbf{\RebelRig},
which is a low-cost CPU-GPU co-processor designed to minimize the cost of a multi-GPU platform.
\Cref{fig:RebelRig_Architecture} visualizes \RebelRig.
It has 4 Nvidia RTX 2070 at 1.7GHz with 8GB memory each,
an Intel i7 6700k desktop processor with 4 cores (8 logical cores with hyperthreading) at 4 GHz with 16GB memory,
and a low-cost crypto mining rig motherboard ASRock H110 Pro BTC+ \cite{ASRockMotherboard}.
The storage device, which keeps the datasets, model checkpoints, and OS (Ubuntu 18.04 LTS),
is a Micron M600 512GB SATA SSD connected through SATA3.

The limitation of this platform is mainly twofold.
(1) It has significantly smaller CPU and GPU memory compared to the other platforms.
(2) Three of the GPUs are connected using PCIe 2.0 x1 to the CPU, while the other one is connected using PCIe 3.0 x16.
This creates an asymmetry across the GPUs (as shown in \Cref{fig:GPUandCPUUtilReb}).

This particular platform is built by us.
Our goal was to build a low-cost platform inspired by cryptocurrency mining specifically repurposed for machine learning,
and understand the relative effectiveness of such a hardware platform in comparison to
more expensive hardware utilized by public cloud providers.

\subsubsection{\SIM}
\label{sec:ExperimentalMethodology:hw:sim}

The second hardware platform is \textbf{\SIM}, which \Cref{fig:Sim_Architecture} visualizes.
It has 4 Nvidia GTX 1080 Ti with 11GB memory each and
two 6-core (12 logical cores with hyperthreading) Intel Xeon E5-2630 v2 processors clocked at 2.6 GHz and 128GB RAM in total.
Each processor has two GPUs attached via PCIe 3.0 x16.
The OS is Centos 7 and installed on a locally-attached HDD,
whereas the rest of the storage needs (the model checkpoints and datasets) are handled via shared storage connected through 1G ethernet.
This increases the startup time by a few seconds when a previous model has to be loaded,
but during active training phase the storage is able to keep up with the processing without being bottlenecked by the storage.

\SIM~platform resembles more what one would find as part of commodity cloud offerings.
However, the hardware components are slightly older.
The CPUs are based on the Ivy Bridge microarchitecture, while the GPUs are based on the Pascal microarchitecture.
Therefore, one can view \SIM~as representative of the previous generation's high-end commodity co-processor.

\subsubsection{\HPC}
\label{sec:ExperimentalMethodology:hw:hpc}

The third hardware platform is \textbf{\HPC}, which \Cref{fig:HPC_Architecture} visualizes.
It has 2 Nvidia Tesla V100 GPUs with 32GB memory each and
two 12-core (24 logical cores with hyperthreading) Intel Xeon Gold 6136 processors at 3.0 GHz and 192GB RAM in total.
The storage is split between a locally-attached SSD and shared storage connected via InfiniBand.
Similarly to \SIM, The OS is Centos 7 and is on locally-attached SSD,
and the rest of the storage needs (the model checkpoints and datasets) are kept on shared storage.
Unlike \SIM, the remote storage does not cause a slow down during initialization thanks to InfiniBand.

\HPC~is composed of the newest microarchitecture technology,
i.e., Intel's Skylake microarchitecture and Nvidia's Volta microarchitecture. 
Therefore, it represents the modern high-end co-processor.

\subsection{Data}
\label{sec:ExperimentalMethodology:Training_Data}

We use the LibriSpeech dataset \cite{LibriSpeech, LibriSpeechPaper}, which falls toward the right-hand side of \Cref{fig:SpeechProperties}.
It is a difficult dataset because it contains continuous speech that uses a large vocabulary with different styles of speech and a large number of speakers.
It is a dataset created from the LibriVox project, which is a collection of audiobooks read by different speakers \cite{LibriVox}. 
LibriSpeech contains 1000 hours of speech and is split into multiple parts.
The \textit{training} data is split into three parts:
100 and 360 hours of \textit{clean} speech, and 500 hours of \textit{other} speech that is more noisy.
In this paper, we train on the combination of these three parts.
Both the \textit{validation} and \textit{test} sets are split into two parts, a \textit{clean} and \textit{other} set,
with each combination containing around 5 hours speech.

For all experiments, we remove all elements above 16700 ms,
since the time duration of the files greatly impacts the time it takes for an inference. 
This results in usage of 95\% of the validation and test data, and 99\% of the training data.
This decision is reasonable since it is not the aim to compete in accuracy but to have enough data to verify generalization of the models.

\subsection{Training}
\label{sec:ExperimentalMethodology:training}

This section details our default training setup,
which is the basis for the experiments in \Cref{sec:results}.

\subsubsection{Framework and Parameters}

The training of the acoustic model is done using TensorFlow version 1.14
following TensorFlow's guides for efficient training
\cite{TF_Data_Input_Pipeline_Performance}.
First, we extract features from the original sound files and store them in compressed TFRecord files before training
\cite{TF_TFRecord}.
Each file contains 64 training samples of Log-Mel and MFCC features and their corresponding text labels.
Extracting the features into intermediate files reduces the number of redundant computations while training thereby shortening the training time.

The movement of data is based on TensorFlow's abstraction called \texttt{tf.data.Dataset}.
This abstraction creates parallel readers that read records from local disk or shared storage.
The readers shuffle, batch and pad elements to supply a buffer.
The elements are padded to a fixed length of 1670 samples, 
which gives a maximum training sample sound length of 16.7 seconds.
The loader fills a buffer containing batches ready to train on. 

The parameters for batch and buffer sizes as well as the number of readers are set to the identified optimal parameters based on preliminary experiments, which \Cref{tab:parameterDiff} displays.
The buffer and batch sizes are different for the different platforms
because of the varying memory capacities.
The main memory consumption during model training increases proportionally
to the batch and buffer sizes.
Therefore, this limits the values for batch and buffer sizes on \RebelRig,
which has smaller memory both on CPU and GPUs.
The number of parallel readers 
was increased until there was no throughput improvements.
Increasing the number of parallel readers to a higher value
than the number of logical threads on \RebelRig~leads to diminishing returns.
On the other platforms, the number of readers do not improve throughput when increased beyond 16 threads.
The reason for this is that the buffer is filled faster than the model is able to process the elements when increasing the number of readers beyond 16. 

\begin{table}
  \centering
  \begin{tabular}{@{ \ }l|rrrr@{ \ }}
  \toprule
  \textbf{Platform}         & \textbf{Total} & \textbf{Batch} & \textbf{Readers}   & \textbf{Buffer} \\
           & \textbf{Batch} & \textbf{per GPU} &  & \\
  \midrule
  \RebelRig     & 96 &     24 &        8  &     40 \\
  \SIM          & 240 &     60 &       16  &    100 \\
  \HPC          & 300 &    150 &       16  &    100 \\
  \bottomrule
  \end{tabular}
  \caption{Default parameters for batch size (number of elements), number of readers, and buffer size (number of batches) for each hardware platform.}
  \label{tab:parameterDiff}
\end{table}

The training was conducted using Synchronous Stochastic Gradient Descent (S-SGD) similar to the related work
\cite{DeepSpeech1, DeepSpeech2, ListenAttendSpell, uniLAS, attention-based-ASR} 
using the AdaDelta optimizer \cite{adadelta}.

Each element in the buffer is an entire batch comprised of multiple elements.
The batch is split into subsets depending on the number of GPUs in use to exploit data parallelism, e.g. a batch of ten elements is split into two subsets of five elements when using two GPUs.
The subsets are each an input to a distinct GPU that contains a copy of the acoustic model described in \Cref{sec:background:AM}.

The model parameters are located on the first GPU indexed as 0,
from which they are also updated and distributed.
Due to the main memory constraints on \RebelRig,
it is better to allocate the model's parameters directly on a GPU
rather than in main memory of the CPU.
This is fair to all platforms since all of them have similar PCIe 3 16x connection to the first GPU. 
This also enables better utilization of the potential direct connections among the GPUs on the platforms instead of necessitating communication with the CPU. 

For each experiment, the batch size is estimated and set to a value that maximizes memory usage on the GPUs
(see \Cref{tab:parameterDiff}).
This maximization of memory is not done on \RebelRig~because increasing the batch size to the memory limit of $160$ reduces the throughput. 
When this value is reduced to $96$ a higher throughput is achieved.
Each training step requires a distribution of the model parameters to the GPUs, which slows down the throughput.
Large batch sizes help to reduce the amount of time used for distributing the model parameters since it changes the relative amount of model parameters and training data \cite[p. 118]{ASRBook}.
However, there is a regularizing effect of having small batches that improves test accuracy.

The model parameters are with floating-point 32-bit precision,
but it is possible for both \RebelRig~and \HPC, to take advantage of 16-bit precision because of their newer GPU microarchitecture.
Using 16-bit or mixed precision \cite{MixedPrecisionTrainingBaidu} increase arithmetic operation efficiency and reduce memory footprint on the GPUs, thereby reducing the training time.
This is not done because \SIM~does not support this optimization, 
and would default to 32-bit precision.
Further work would be to explore the gains of 16-bit precision, and potentially going even further with quantized neural networks \cite{QNN} that utilize 8-bit values which have recently shown great promise without significant loss in accuracies \cite{xilinxNNBench}.

\subsubsection{The States of Training}

\begin{figure}
  \centering
  \includegraphics[width=\linewidth]{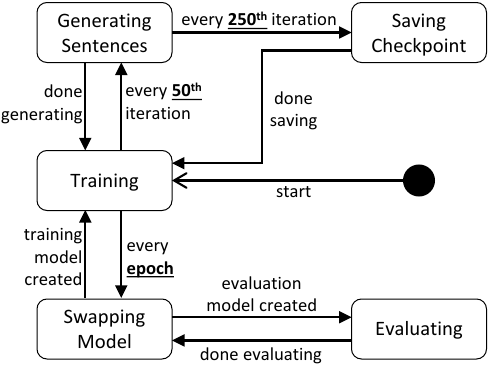}
  \caption{States during training.}
  \label{fig:TrainingStateMachine}
\end{figure}

The training is conducted while monitoring loss, CER, WER, and output sentences.
Calculating the loss, CER, and WER
(as defined in \Cref{sec:background:AM:ctcbeam} and \Cref{sec:ExperimentalMethodology:metrics:accuracy})
requires evaluation of the model using the validation set.
This means that the training and evaluation needs to happen simultaneously. 

\Cref{fig:TrainingStateMachine} illustrates the training process in the form of a state machine.
The transition between the states includes the iteration interval that triggers the state transition. 
The initial state is \emph{training}, where the model is fitted to the training data by continuously iterating through different batches. 
Once the training iterates through 50 batches,
we switch to a new state that \emph{generates sentences} from the training dataset based on the current model with a beam width of one. 
Every 250th batch iteration, a \emph{checkpoint} of the model is saved.
The checkpoint is used when running evaluation.
It can also be used to restart the training in case the training has been interrupted.
The model is \textit{evaluated} every epoch by running the model on the validation set.
To do this, the training model needs to be removed from the GPUs and the evaluation model needs to be loaded.
\emph{Swapping Model} in \Cref{fig:TrainingStateMachine} represents this action.
Swapping the model has significant influence on the throughput, and is further analyzed in \Cref{sec:results}. 

There are alternatives to avoid model swapping during the training process.
An alternative is to always allocate memory on the GPUs for a process that evaluates the model whenever a checkpoint is saved
(as adopted by Mozilla Deep Speech implementation \cite{DeepSpeechGithub}).
The drawback of this alternative approach is that allocating memory for the evaluation limits the available memory for the model training. 
This is especially an obstacle when training on a platform like \RebelRig~because the memory is already quite limited. 
Another alternative is to avoid the interleaving of the training and evaluation on the same platform by training and evaluating on separate machines. 
This would require more complex and expensive hardware setup
in addition to the more complicated management of the dataflow across machines. 

For the evaluation phase,
there is a tradeoff between the beam width selected for decoding the resulting sentence (\Cref{sec:background:AM:ctcbeam})
and the accuracy achieved (\Cref{sec:ExperimentalMethodology:metrics:accuracy}).
\Cref{tab:BeamWidthTimings} shows how the evaluation time and accuracy change as the beam width increases,
based on measurements done using a single GPU on \HPC.
Based on these results, the beam width is set to 256 while training to reduce the evaluation time.
For the final validation once training is stopped, the beam width is also set to 256 for consistency.

\begin{table}
    \centering
    \begin{tabular}{@{ \ }rrll@{ \ }}
    \toprule
    BW        & Time (sec) & WER   & CER   \\
    \midrule
    64        & 154  & 25.20 & 9.315 \\
    128       & 187  & 25.02 & 9.179 \\
    256       & 266  & 24.95 & 9.069 \\
    512       & 443  & 24.78 & 9.029 \\
    1024      & 849  & 24.75 & 8.993 \\
    2048      & 1860 & 24.68 & 8.968 \\
    4096      & 4200 & 24.62 & 8.943 \\
    \bottomrule
    \end{tabular}
    \caption{Evaluation time in seconds, WER, and CER for different beam width values (BW)}
    \label{tab:BeamWidthTimings}
\end{table}

\subsubsection{The Model}

The model has the same parameters for the different platforms except the differences mentioned in \Cref{tab:parameterDiff}.
The model is smaller and simpler compared to related work (\Cref{sec:related_work:nn}),
but still contains the same building blocks and overall topology (\Cref{fig:model}).

The input features for the model is the concatenation of the Log Mel spectrogram of 80 dimensions and the 13 dimensions of MFCC, giving a 93-dimensional input feature per 10 ms time-step. 
The CNN has a kernel size of 11 x 93 x 600 which means it takes all the input features of 11 consecutive time-steps, and produce a 600-dimensional output feature.
The CNN is applied with a stride of 2, and the CNN is only applied to valid inputs.
This means that the model cannot take inputs shorter than 11 time-steps.
The application of stride and valid requirements reduce the number of time-steps in the output of the CNN according to the following formula assuming that $x$ is the input size in the time dimension, $y$ is the output size after CNN, $c_w$ is convolution width that is 11 in our case, $c_s$ is the stride and $x > c_w$.

\begin{equation}
   y = \ceil*{\frac{x - c_w}{c_s}} 
\end{equation}

Following the CNN is the LSTMs that have size 800. The first LSTM layer takes the 600-dimensional input and all subsequent layers take the 800-dimensional output from the previous LSTM layer.
The output of the fifth layer is passed to a feed-forward layer that transforms the output to character probability distributions of dimension 30, which represent \{space, a, b, ... , z, ', separation character, blank character\}.
The separation character is used in cases where the word contains double letters, such as the word "cool", having the output "co\_ol", where "\_" is used to represent the separation character. 
Between each layer, batch normalization and a dropout of 5\% are applied.

\section{Results}
\label{sec:results}

\newcommand{\TTAtable}{
    \begin{table}
        \centering
        \begin{tabular}{@{ \ }l|rrr|rrr@{ }}
            \toprule
            \multirow{2}{*}{\diagbox[width=5em]{\textbf{Sys}}{\textbf{TTA}}}          & \multicolumn{3}{c|}{\textit{CER}}   & \multicolumn{3}{c}{\textit{WER}}    \\
                                                                        & \multicolumn{1}{c}{\textbf{10.0 } }    & \multicolumn{1}{c}{\textbf{8.0}}     & \multicolumn{1}{c|}{\textbf{7.0}}    & \multicolumn{1}{c}{\textbf{22.0}}      & \multicolumn{1}{c}{\textbf{20.2}}     & \multicolumn{1}{c}{\textbf{19.2}} \\
            \midrule          
            \RebelRig    &   820      &  1641    & 2450    &  1908    &  2174    & 2725     \\
                         &   100\%    &   100\%  &  100\%  &   100\%  &   100\%  &  100\%   \\
            \hdashline             
            \SIM         &   672      &  1118    & 2429    &  1251    &  1980    & 2834     \\
                         &    82\%    &    62\%  &   99\%  &    66\%  &    91\%  &  104\%   \\
            \hdashline
            \HPC         &   511      &   974    &         &  1143    &  2840    &          \\ 
                         &    62\%    &    59\%  &   NA    &    59\%  &   131\%  &   NA     \\ 
            \bottomrule
            \end{tabular}
        \caption{TTA values expressed in minutes and relative time to \RebelRig~for WER and CER of the different platforms.}
        \label{tab:TTA}
    \end{table}
}

\newcommand{\EndPerformTable}{
\begin{table}
    \centering
    \begin{tabular}{@{ \ }l|rr| rr@{ \ }}
        \toprule
                    &  \multicolumn{2}{c|}{\textit{Clean}}  &  \multicolumn{2}{c}{\textit{Other}}     \\ 
        \textbf{System}      &  \textbf{Val}           & \textbf{Test}       &  \textbf{Val}           & \textbf{Test}          \\
        \midrule    
        \RebelRig   &  17.90         &  17.32     &     41.85      &   45.04   \\
        \SIM        &  18.50         &  18.68     &     44.31      &   48.15   \\
        \HPC        &  19.86         &  19.45     &     45.86      &   49.43   \\ 
        \midrule
        \multicolumn{5}{l}{\textbf{Related Work}}\\
        \midrule
        Deep Speech 2 \cite{DeepSpeech2} 
                    & NA             &   5.15     & NA             &   12.73  \\
        Lüscher et al. \cite{InterSpeechStateOfTheArtPerformanceLibrispeech}
                    &   1.90         &   2.30     &     4.50       &    5.00    \\
        \bottomrule
        \end{tabular}
    \caption{Final WER of trained model with beam width 512.}
    \label{tab:EndPerformance}
\end{table}
}

\newcommand{\LossCERVERFigures}{

\begin{figure*}
    \begin{subfigure}{0.33\linewidth}
        \centering
        \includegraphics[width=\linewidth]{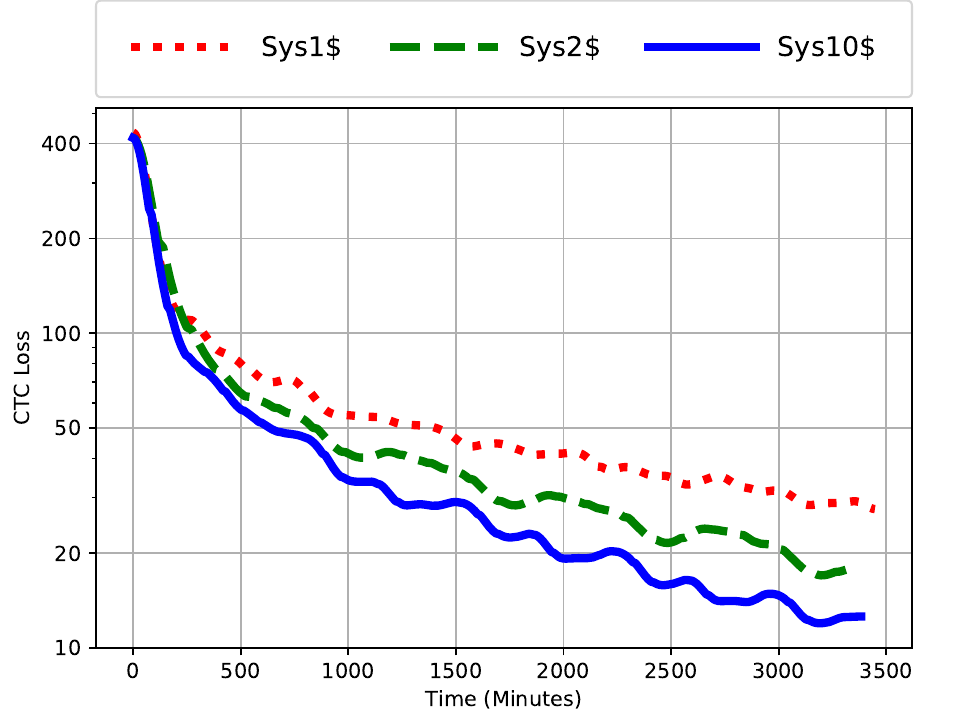}
        \captionsetup{width=.9\linewidth}
        \subcaption{Smoothed loss of training set over time with logarithmic y-axis.}
        \label{fig:LossOverTime}
    \end{subfigure}
    \begin{subfigure}{0.33\linewidth}
        \centering
        \includegraphics[width=\linewidth]{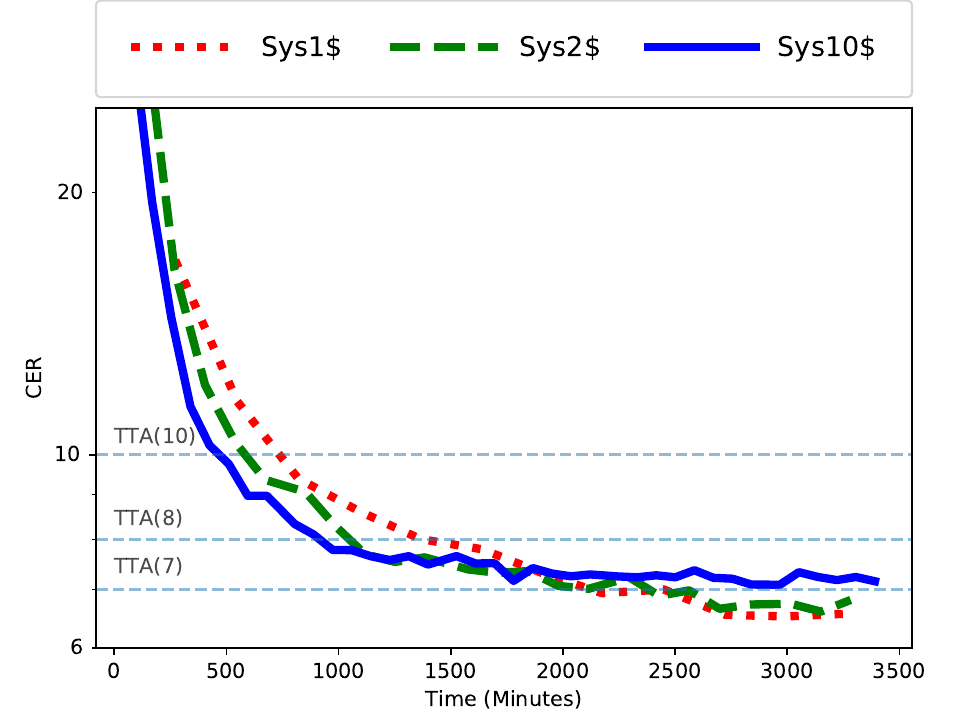}
        \captionsetup{width=.9\linewidth}
        \subcaption{CER of validation set over time with logarithmic y-axis.}
        \label{fig:CERdevset}
    \end{subfigure}
    \begin{subfigure}{0.33\linewidth}
        \centering
        \includegraphics[width=\linewidth]{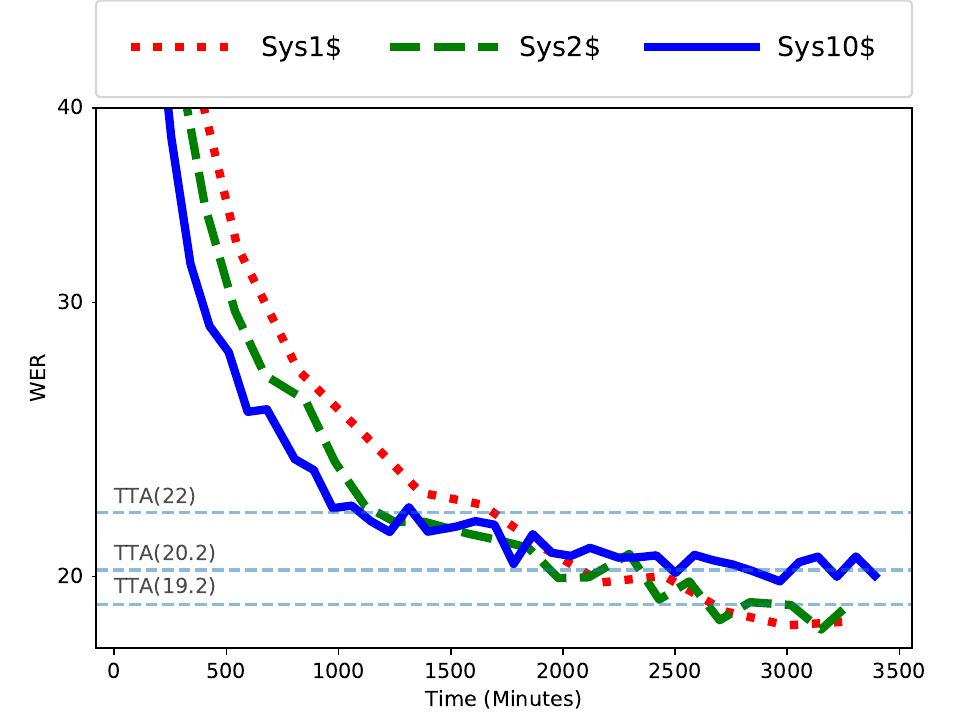}
        \captionsetup{width=.9\linewidth}
        \subcaption{WER of validation set over time with logarithmic y-axis.}
        \label{fig:WERdevset}
    \end{subfigure}
    \caption{Loss, Character Error Rate (CER), and Word Error Rate (WER) over time to measure Time-To-Accuracy (TTA).}
    \label{fig:accuracies}
\end{figure*}

}

\newcommand{\UtilizationFigures}{

\begin{figure*}
    \begin{subfigure}{0.33\linewidth}
        \centering
        \includegraphics[width=\linewidth]{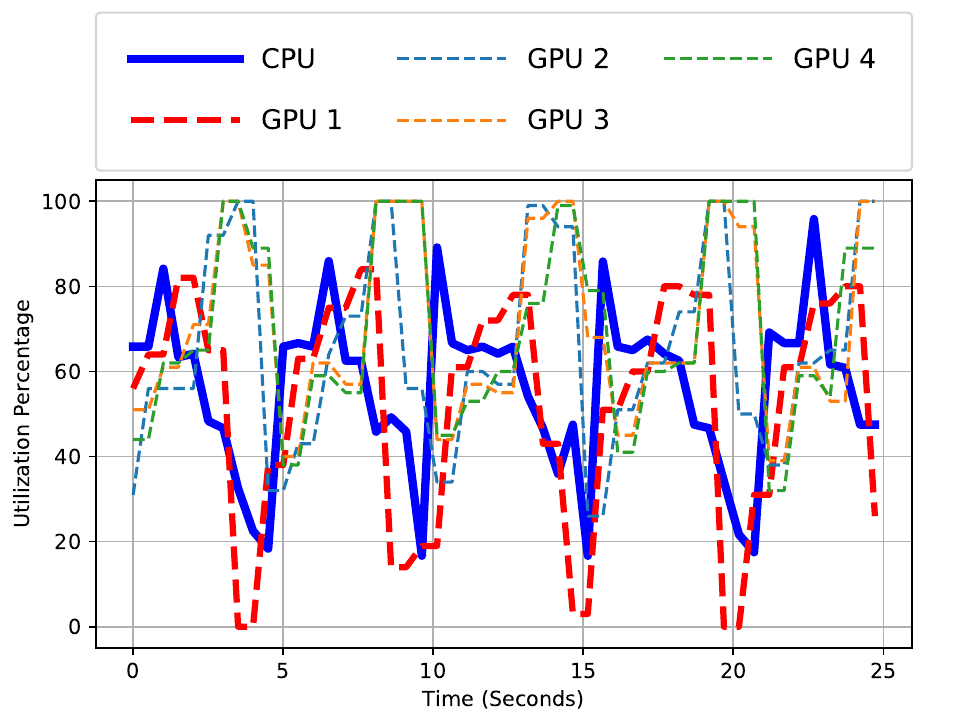}
        \caption{\RebelRig}
        \label{fig:GPUandCPUUtilReb}
        \captionsetup{width=.9\linewidth}
    \end{subfigure}
    \begin{subfigure}{0.33\linewidth}
        \centering
        \includegraphics[width=\linewidth]{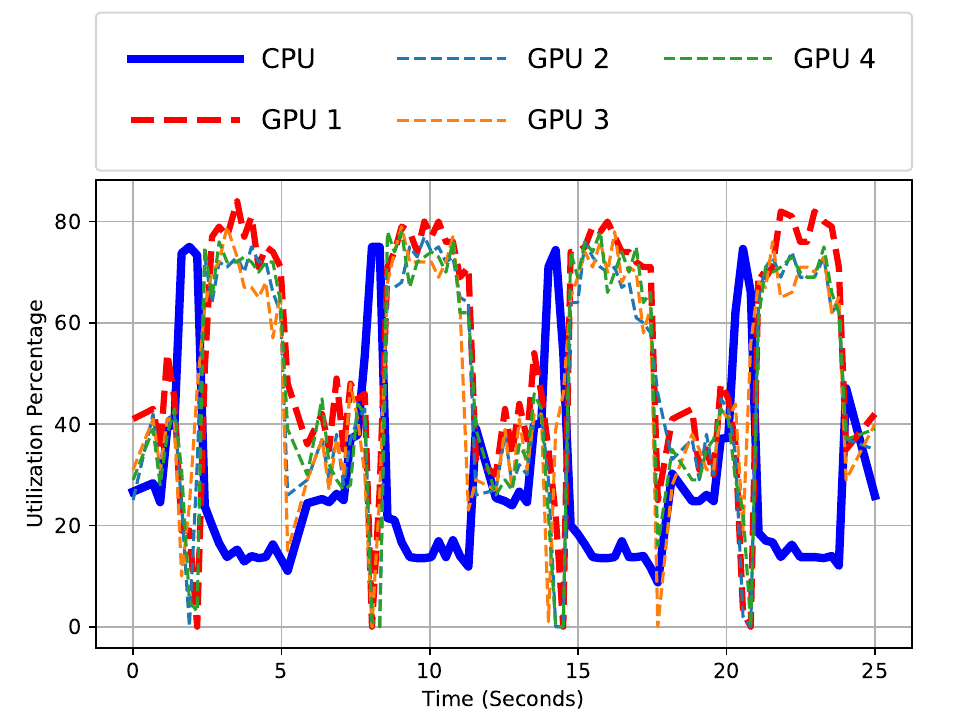}
        \caption{\SIM}
        \label{fig:GPUandCPUUtilSim}
        \captionsetup{width=.9\linewidth}
    \end{subfigure}
    \begin{subfigure}{0.33\linewidth}
        \centering
        \includegraphics[width=\linewidth]{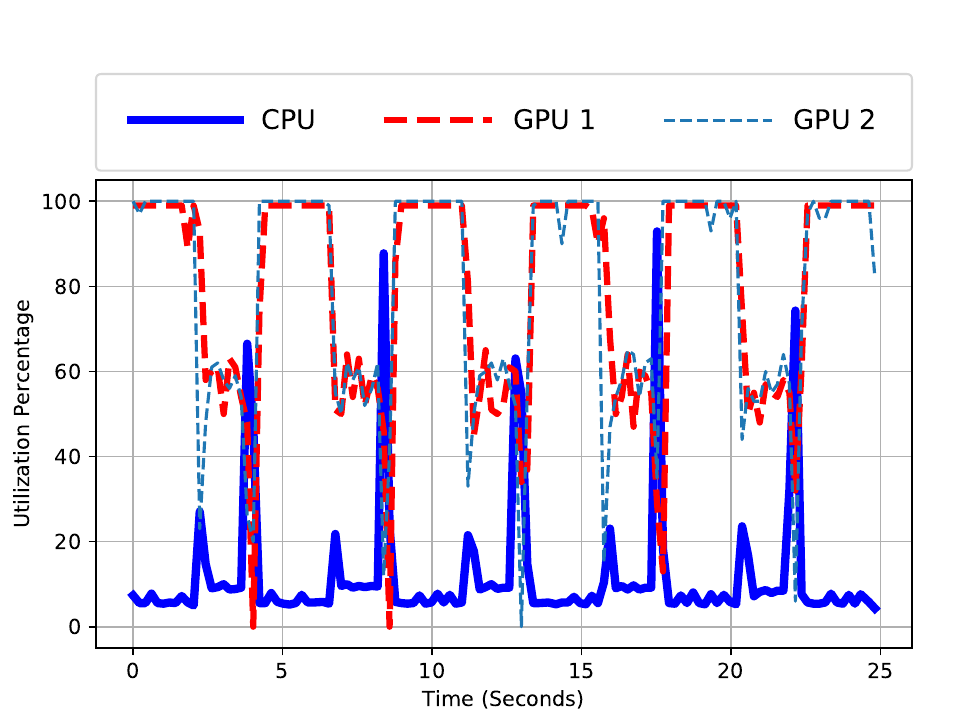}
        \caption{\HPC}
        \label{fig:GPUandCPUUtilHPC}
        \captionsetup{width=.9\linewidth}
    \end{subfigure}
    \caption{Hardware utilization over time on three co-processors. The figure focuses on a 25 second interval for ease of visualization. The utilization trends are stable over time.}
    \label{fig:Utilization}
\end{figure*}
}

\newcommand{\ThroughputFigure} {
    \begin{figure}
        \centering
        \includegraphics[width=\linewidth]{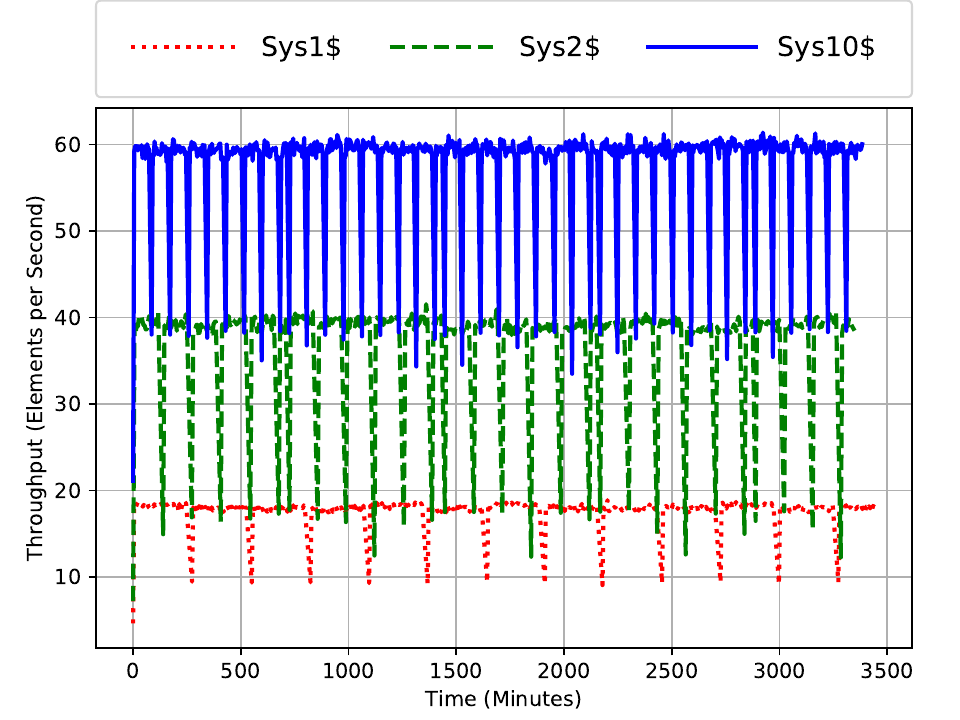}
        \caption{Processing throughput during training.}
        \label{fig:Throughput}
    \end{figure}
}

\newcommand{\ThroughputFigureSmallBatches} {
    \begin{figure}
        \centering
        \includegraphics[width=\linewidth]{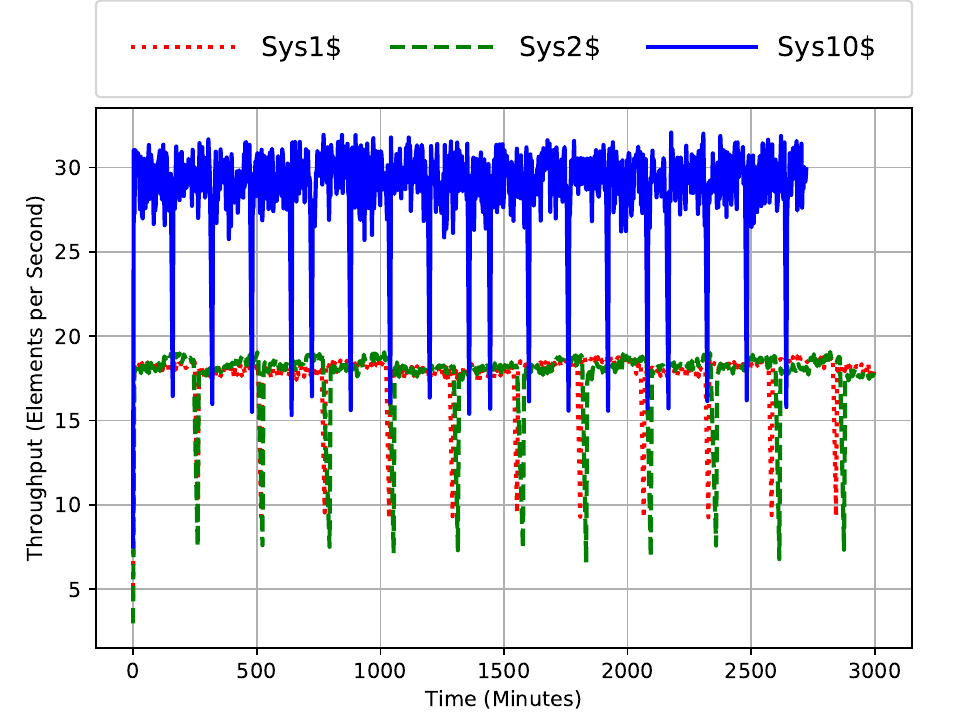}
        \caption{Processing throughput during training with equal batch sizes.}
        \label{fig:ThroughputLowerBatchsize}
    \end{figure}
}

\newcommand{\TTAtableSmallBatches}{
    \begin{table}
        \centering
        \begin{tabular}{@{ \ }l|rrr|rrr@{ }}
            \toprule
            \multirow{2}{*}{\diagbox[width=5em]{\textbf{Sys}}{\textbf{TTA}}}          
            & \multicolumn{3}{c|}{\textit{CER}}   & \multicolumn{3}{c}{\textit{WER}}    \\
                & \multicolumn{1}{c}{\textbf{10.0 } }  & \multicolumn{1}{c}{\textbf{8.0}}    & \multicolumn{1}{c|}{\textbf{7.0}}    
                & \multicolumn{1}{c}{\textbf{22.0}}      & \multicolumn{1}{c}{\textbf{20.0}}     & \multicolumn{1}{c}{\textbf{19.0}} \\
            \midrule          
            \RebelRig    &   771      &  1291    & 2322    &  1806    &  2322    & 2322     \\
                         &   100\%    &   100\%  &  100\%  &   100\%  &   100\%  &  100\%   \\
            \hdashline             
            \SIM         &   791      &  1571    & 2091    &  1571    &  2091    & 2352     \\
                         &   103\%    &   122\%  &   90\%  &    87\%  &    90\%  &  101\%   \\
            \hdashline
            \HPC         &   479      &   879    & 1600    &  1197    &  1600    & 2078     \\ 
                         &    62\%    &    68\%  &   69\%  &    66\%  &    68\%  &   89\%     \\ 
            \bottomrule
            \end{tabular}
        \caption{TTA values expressed in minutes and relative time to \RebelRig~for WER and CER of the different platforms with reduced batch sizes.}
        \label{tab:TTALowerBatchsize}
    \end{table}
}

This section splits the results into three parts:
(1) loss values, accuracy, and TTA,
(2) hardware utilization and throughput, and
(3) a comparison of the platforms with the same batch size.

\LossCERVERFigures
\TTAtable

\subsection{Loss, Accuracy, \& TTA}
\label{sec:results:tta}

\Cref{fig:LossOverTime} has the CTC \textbf{loss values} over time for the three hardware setups. The CTC loss axis is logarithmic to better represent the nuances over time. 
The loss values have been smoothed by applying a 1-dimensional Gaussian filter from SciPy with a kernel standard deviation of 10 because the actual loss value oscillates throughout training. 
Looking at \Cref{fig:LossOverTime}, it can be observed that on all hardware setups, the training loss is reduced over time which means that the model is able to adapt to the training data.
\HPC~has the fastest convergence, \SIM~has the second fastest convergence, and \RebelRig~has the slowest convergence. 
This result corresponds to the costs of the three platforms with \HPC~having the highest cost and \RebelRig~having the lowest cost.
All three executions stay above a loss value of 10 and they are stopped within approximately 3400 minutes which is equivalent to two days and 8 hours. 

The \textbf{accuracy} of the clean validation set over time is shown in \Cref{fig:CERdevset} and \Cref{fig:WERdevset}. 
The figures also have a logarithmic y-axis to express the nuances of the accuracy through all time steps. 
The TTA levels chosen for CER are 10, 8, and 7. The TTA levels chosen for WER are 22, 20.2, and 19.2. 
All TTA levels have been plotted as horizontal lines to indicate when the different executions reach the TTA levels. 
\Cref{tab:TTA} reports the specific TTA values, where $NA$ represents the case,
where the specified accuracy level is not reached.

The initial 1000 minutes of the CER has a development similar to the training loss, where \HPC~converges quicker than \SIM, which in turn converges quicker than \RebelRig. 
This situation is in effect at both TTA(10) and TTA(8), but it is subsequently turned upside down. 
\SIM~surpasses \HPC~, and it is later surpassed by \RebelRig. 
\SIM~and \RebelRig~reache TTA(7) approximately at the same time step,
while \HPC~never reaches this CER, which is denoted by $NA$ in \Cref{tab:TTA}.  
This demonstrates the effect of large batch sizes. The large batch size of \HPC~improves the pace of the convergence initially, but ends up impeding the accuracy of the model. 
Similar results of reduced statistical efficiency with larger batch sizes were also identified in \cite{Crossbow}.
The reduced statistical efficiency stems from the higher generalization error of models trained with large batch sizes, which is a problem known for many years \cite{DeepLearningGoodfellow}.

\Cref{fig:WERdevset} has a similar development with TTA(20) being reached first by \HPC~and TTA(20.2) and TTA(19.2) reached by the two other co-processors first. 
The general level of WER is higher than CER, but the relative accuracies of the different platforms are similar.

\EndPerformTable

The accuracies described above are calculated on the validation dataset, but when reporting the actual accuracy of the model, it has to be calculated on the separate test dataset. 
\Cref{tab:EndPerformance} reports the accuracies with the test dataset.
The evaluation is done on both the \emph{clean} and the \emph{other} dataset and the table also includes the accuracy results of the validation set, which is referred to as \emph{Val}. 
The accuracies reported in related work is also included in the table for comparison. 
We include the results from Deep Speech 2 \cite{DeepSpeech2}, because the model is similar to our work, and the results from Lüscher et al. \cite{InterSpeechStateOfTheArtPerformanceLibrispeech}, because it represents the state-of-the-art accuracy for this specific dataset. 

The model in our work is similar to Deep Speech 2, but it diverges in two ways which results in shorter training time and lower accuracy: 
(1) CNN output size is 600 in our system and 1280 in Deep Speech 2, and
(2) LSTM size is 800 in our system and 1510 in Deep Speech 2.
By reducing the sizes of the model layers, accuracy is traded for faster computation. 

Quoting Baidu \cite{DeepSpeech2} commenting on the model size of Deep Speech 2: "Training a single model at these scales requires tens of exaFLOPs that would require 3-6 weeks to execute on a single GPU."
This was in late 2015 and their model was trained on 8-16 unspecified GPUs that are older GPU versions than the Nvidia Tesla V100 GPUs used by \HPC~in this project. 

The accuracy of Lüscher et al. \cite{InterSpeechStateOfTheArtPerformanceLibrispeech} is better than our accuracy and even better than Deep Speech 2.
The difference is that Lüscher et al. base their model on the attention-based architecture, which is also found in \cite{ListenAttendSpell} and originates from \cite{originalAttentionModel}.
The attention-based model clearly performs better, but it requires longer training time, especially with the model parameters used for their specific solution. 
A scaled down version of an attention-based architecture would also have been appropriate for this paper.

\UtilizationFigures

\subsection{Utilization \& Throughput}
\label{sec:results:util}

\Cref{fig:Utilization} shows the \textbf{utilization} of the three platforms
over a 25 second interval. 
The solid line represents the utilization of the CPU over the last 0.03 seconds, while the dashed lines are the GPUs' utilization.
The horizontal axis is time in seconds, and the vertical axis is the utilization percentage of the corresponding GPU and CPU.
The CPU utilization is the average across all cores for the specific training process.

Looking at the utilization of \HPC~in \Cref{fig:GPUandCPUUtilHPC}, three phases of utilization during a single batch processing can be observed.
The first utilization phase is represented by a spike on the CPU when the training data and model is transferred to the GPU. 
These spikes are most visible in \Cref{fig:GPUandCPUUtilHPC} at 4, 8, and 17 seconds, but are also present at 13 and 23 seconds. 
The second utilization phase happens on the GPUs when they are using the acoustic model to infer results from the input data. 
This is seen in \Cref{fig:GPUandCPUUtilHPC} where the GPUs are capped at 100\% utilization just after the CPU spikes.
The last phase is where the model is collected, gradients are summed, and the model is updated. 
This is seen as a small bump on the CPU utilization after the utilization of the GPUs are reduced.
After completion of the three phases, the entire process is repeated. 

\SIM's utilization is shown in \Cref{fig:GPUandCPUUtilSim}.
The spikes are wider in this graph, indicating a longer startup for each iteration
as expected due to the slower CPU and main memory (as explained in \Cref{sec:ExperimentalMethodology:hw:sim}). 
This gives a clearer view of the 0\% utilization of the GPUs while transferring the data.
This 0\% utilization is also present in \cref{fig:GPUandCPUUtilHPC} but it is less visible in that figure due to the shorter duration.
Looking at \cref{fig:GPUandCPUUtilSim}, it is observed that the GPU utilization is capped at around 75\%. 
The model and batch size use all of the available memory on the GPUs, but the cores are not fully utilized. 
This means that neither the model size nor the batch size can be further increased to fully utilize the cores because the allocated memory would exceed the memory limit. 

The utilization phases appear in both \Cref{fig:GPUandCPUUtilHPC} and \Cref{fig:GPUandCPUUtilSim}, but \Cref{fig:GPUandCPUUtilReb} has a slightly different pattern. 
GPU 1 is connected using PCIe 3 x16 and the rest of the GPUs are connected with PCIe 2 x1. 
GPU 1 is highly utilized in the beginning of each batch iteration and then quickly drops in utilization.
On the other hand, GPU 2, GPU 3, and GPU 4 have moderate utilization before being fully utilized around the time GPU 1 is done with its batch.
This utilization difference is caused by the different connection types between the graphics cards.
When GPU 1 finishes earlier, it waits for the three other GPUs to finish. During this time, the GPU almost goes to 0\% utilization. 

To prevent this low utilization of the GPU with the current heterogenous connections, the training could be based on asynchronous SGD (A-SGD) \cite{NothingReallyMatters} rather than synchronous SGD (S-SGD). 
In S-SGD, all GPUs have a consistent view of the model and each GPU has to wait for all other GPUs to produce their partial gradients so that the partial gradients can be aggregated and the model updated before starting the next iteration. In A-SGD, each GPU immediately starts the next iteration after its partial gradient is added to the accumulated gradient \cite{Crossbow}. 

Another difference in \Cref{fig:GPUandCPUUtilReb} compared to the
other platforms is the CPU utilization.
The CPU has fewer logical cores compared to the CPUs in the other systems. 
This means that the CPU is generally more utilized which gives less visible spikes, but the spikes are still noticeable at 1, 6, 10, 16, and 22 seconds.

\ThroughputFigure

Looking at \Cref{fig:Throughput}, it is seen that the three systems follow the same pattern as in the utilization figures. \HPC~has better utilization and throughput than \SIM~, which has better utilization and throughput than \RebelRig. 
The \textbf{throughput} oscillates greatly, but the minimum and maximum goes to the same levels over time for the respective systems. The low points of the throughput occurs every time the training model is replaced by the evaluation model, which was explained in \Cref{sec:ExperimentalMethodology:training}. 
The throughput is reduced by approximately 30-50\% on all systems during this swap. 

Comparing the utilization and throughput figures to the accuracy development in \Cref{fig:accuracies}, it is evident that the highly utilized, high throughput systems achieves lower accuracies eventually. The throughput and utilization analyses represent the hardware efficiency, whereas \Cref{fig:CERdevset} and \Cref{fig:WERdevset} represent the training efficiency. 
This means that \HPC~has high hardware efficiency but low training efficiency, and \RebelRig~has low hardware efficiency but high training efficiency. 
This demonstrates the point that when training machine learning models, the training efficiency is more important than the hardware efficiency. 

The utilization and throughput analyses and following optimizations are however not purposeless, since they can be used to identify bottlenecks as was done with GPU 1 of \RebelRig~with PCIe 3 connection. This kind of analysis makes it possible to improve the overall efficiency using the current hardware setup by for instance changing from an S-SGD to an A-SGD. 

\subsection{Parameter Equality Analysis}
\label{sec:results:sa}

To confirm that the achieved accuracy from \RebelRig~can be achieved on the other platforms, the parameters used when training on \RebelRig~is applied to the training on \HPC~and \SIM.
The batch size is reduced to the same value with the difference that \HPC~has two GPUs on which the batch is split unlike \RebelRig~and \SIM~that have four. We also chose to reduce the beam width to 64 while evaluating thereby reducing overall training time. 
The change in beam width only reduces accuracy without effecting the convergence.

The reduction in batch size impedes the throughput as seen in \Cref{fig:ThroughputLowerBatchsize}.
It shows that the total throughput is reduced to half on both \HPC~and \SIM~compared to their previous throughput seen in \Cref{fig:Throughput}.
The utilization of the GPUs on HPC is also found to be reduced, going from 76\% utilization to 67\% on average.
It is also clear that \SIM~and \RebelRig~have the same performance in throughput and therefore should converge similarly, which they do, while \HPC~converges faster than the other two systems due to its higher throughput.

\ThroughputFigureSmallBatches

The TTA values achieved by the models with these reduced batch sizes can be seen in \Cref{tab:TTALowerBatchsize}.
The results show that the accuracy of the model is able to reach the same level on \HPC~as on \RebelRig, but with a better TTA on \HPC.

\TTAtableSmallBatches

\section{Summary \& Conclusion}
\label{sec:conclusion}

This paper studied the behavior of the training of an acoustic model for ASR
on different types of CPU-GPU co-processor hardware
that fall into different price categories.
The acoustic model was based on state-of-the-art proposals
that use deep neural networks for this purpose.
Our goal was to observe the impact of higher-end processors
in comparison with lower-end ones in this problem domain.
By focusing on time-to-accuracy as the main metric,
we observed that  utilizing hardware acceleration yields good results
even without high-end equipment.

The low-budget \RebelRig~achieved a WER, CER, and TTA lower than the high-end \HPC, even though the throughput of \RebelRig~was lower. 
This result is caused by the low batch size used in \RebelRig, which improves the training efficiency of the model. 
When using the same batch size on all hardware platforms, the throughput of \RebelRig~and \SIM~are similar, but the throughput of \HPC~is still higher. 
The accuracy of \RebelRig~and \SIM~are similar in this case and \HPC~achieves an even better accuracy, but \HPC~and \SIM~are underutilized.

The latest generations of co-processor hardware, such as the one evaluated as \HPC,
offer a huge computation and acceleration power.
Such co-processors are becoming more and more widely available to the end-users.
The embarrassingly parallel nature of most neural network tasks
make them great for exploiting these types of hardware.
However, there is no free lunch, especially for the more complex application
domains like ASR.
One has to pay attention to design both a statistically accurate model
and a hardware-conscious one based on the available processing units
in order to avoid wasting hardware resources.

Going forward, we have various options to investigate.
There is already well-established frameworks (TensorFlow, Torch, etc.) for crafting neural networks
and work on providing hardware-conscious libraries for data scientists \cite{pynq, rapids}.
We need to invest further in these frameworks and libraries,
and understand their behavior on different types of processing units in more detail.
It is unreasonable to expect every data scientist crafting models
based on neural networks for a specific problem to be hardware gurus.
However, it is also unreasonable to underutilize extremely powerful hardware.
Furthermore,
similarly to the research on co-locating different tasks on cloud,
we should also investigate ways to co-locate different types of model training
to exploit idle-sitting hardware resources
without breaking the performance of individual training processes.


\bibliographystyle{abbrv}

\balance

\end{document}